\documentclass[12pt,aps,epsf,preprint]{article}

\usepackage{amssymb}
\usepackage{amsmath}

\def \be {\begin{equation}}
\def \ee {\end{equation}}
\def \bea {\begin{eqnarray}}
\def \eea {\end{eqnarray}}
\def \nn {\nonumber}

\def \a {\alpha}
\def \b {\beta}
\def \g {\gamma}
\def \G {\Gamma}
\def \d {\delta}

\def \m {\mu}
\def \n {\nu}
\def \k {\kappa}

\def \s {\sigma}
\def \r {\rho}
\def \o {\omega}
\def \O {\Omega}
\def \th {\theta}
\def \Th {\Theta}

\def \t {\tau}
\def \dag {\dagger}
\def \p {\partial}

\def\bd{\begin{document}}
\def\ed{\end{document}}
\def\nn{\nonumber}
\def\bea{\begin{eqnarray}}
\def\eea{\end{eqnarray}}
\let\bm=\bibitem
\let\la=\label

\def\N{{\cal N}}
\def\sst{\scriptscriptstyle}
\def\thetabar{\bar\theta}
\def\Tr{{\rm Tr}}
\def\one{\mbox{1 \kern-.59em {\rm l}}}

\def\a{\alpha}      \def\da{{\dot\alpha}}
\def\b{\beta}       \def\db{{\dot\beta}}
\def\c{\gamma}  \def\C{\Gamma}  \def\cdt{\dot\gamma}
\def\d{\delta}  \def\D{\Delta}  \def\ddt{\dot\delta}
\def\e{\epsilon}        \def\vare{\varepsilon}
\def\f{\phi}    \def\F{\Phi}    \def\vvf{\f}
\def\h{\eta}
\def\k{\kappa}
\def\l{\lambda} \def\L{\Lambda}
\def\m{\mu} \def\n{\nu}
\def\o{\omega}
\def\P{\Pi}
\def\r{\rho}
\def\s{\sigma}  \def\S{\Sigma}
\def\t{\tau}
\def\th{\theta} \def\Th{\Theta} \def\vth{\vartheta}
\def\X{\Xeta}
\def\z{\zeta}
\def\w{\wedge}
\def\u{\underline}

\def\cA{{\cal A}} \def\cB{{\cal B}} \def\cC{{\cal C}}
\def\cD{{\cal D}} \def\cE{{\cal E}} \def\cF{{\cal F}}
\def\cG{{\cal G}} \def\cH{{\cal H}} \def\cI{{\cal I}}
\def\cJ{{\cal J}} \def\cK{{\cal K}} \def\cL{{\cal L}}
\def\cM{{\cal M}} \def\cN{{\cal N}} \def\cO{{\cal O}}
\def\cP{{\cal P}} \def\cQ{{\cal Q}} \def\cR{{\cal R}}
\def\cS{{\cal S}} \def\cT{{\cal T}} \def\cU{{\cal U}}
\def\cV{{\cal V}} \def\cW{{\cal W}} \def\cX{{\cal X}}
\def\cY{{\cal Y}} \def\cZ{{\cal Z}}

\def\ua{\underline{\alpha}}
\def\ub{\underline{\phantom{\alpha}}\!\!\!\beta}
\def\uc{\underline{\phantom{\alpha}}\!\!\!\gamma}
\def\um{\underline{\mu}}
\def\ud{\underline\delta}
\def\ue{\underline\epsilon}
\def\una{\underline a}\def\unA{\underline A}
\def\unb{\underline b}\def\unB{\underline B}
\def\unc{\underline c}\def\unC{\underline C}
\def\und{\underline d}\def\unD{\underline D}
\def\une{\underline e}\def\unE{\underline E}
\def\unf{\underline{\phantom{e}}\!\!\!\! f}\def\unF{\underline F}
\def\unm{\underline m}\def\unM{\underline M}
\def\unn{\underline n}\def\unN{\underline N}
\def\unp{\underline{\phantom{a}}\!\!\! p}\def\unP{\underline P}
\def\unq{\underline{\phantom{a}}\!\!\! q}
\def\unQ{\underline{\phantom{A}}\!\!\!\! Q}
\def\unH{\underline{H}}

\def\As {{A \hspace{-6.4pt} \slash}\;}
\def\bs {{b \hspace{-6.4pt} \slash}\;}
\def\Ds {{D \hspace{-6.4pt} \slash}\;}
\def\ds {{\del \hspace{-6.4pt} \slash}\;}
\def\ss {{\s \hspace{-6.4pt} \slash}\;}
\def\ks {{ k \hspace{-6.4pt} \slash}\;}
\def\ps {{p \hspace{-6.4pt} \slash}\;}
\def\pas {{{p_1} \hspace{-6.4pt} \slash}\;}
\def\pbs {{{p_2} \hspace{-6.4pt} \slash}\;}

\def\Fh{\hat{F}}
\def\Vh{\hat{V}}
\def\Xh{\hat{X}}
\def\ah{\hat{a}}
\def\xh{\hat{x}}
\def\yh{\hat{y}}
\def\ph{\hat{p}}
\def\xih{\hat{\xi}}

\def\psit{\tilde{\psi}}
\def\Psit{\tilde{\Psi}}
\def\tht{\tilde{\th}}

\def\At{\tilde{A}}
\def\Qt{\tilde{Q}}
\def\Rt{\tilde{R}}
\def\Nt{\tilde{N}}

\def\at{\tilde{a}}
\def\st{\tilde{s}}
\def\ft{\tilde{f}}
\def\pt{\tilde{p}}
\def\qt{\tilde{q}}
\def\vt{\tilde{v}}
\def\nt{\tilde{n}}

\def\delb{\bar{\partial}}
\def\bz{\bar{z}}
\def\bD{\bar{D}}
\def\bB{\bar{B}}

\def\bk{{\bf k}}
\def\bl{{\bf l}}
\def\bp{{\bf p}}
\def\bq{{\bf q}}
\def\br{{\bf r}}
\def\bx{{\bf x}}
\def\by{{\bf y}}
\def\bR{{\bf R}}
\def\bV{{\bf V}}

\def\d{\delta}\def\D{\Delta}\def\ddt{\dot\delta}

\def\p{\partial} \def\del{\partial}
\def\xx{\times}
\def\uno{\mbox{1 \kern-.59em {\rm l}}}

\def\trp{^{\top}}
\def\inv{^{-1}}
\def\dag{{^{\dagger}}}
\def\pr{^{\prime}}

\def\rar{\rightarrow}
\def\lar{\leftarrow}
\def\lrar{\leftrightarrow}

\begin{document}
{\hfill {CAS-KITPC/ITP-013}} \vspace{2cm}
\begin{center}
{\Large The Self-dual String Soliton in  $AdS_4\times S^7$
spacetime }
\vspace{2cm}\\[10mm]
{\large Bin Chen\footnote{Email:bchen01@pku.edu.cn}}

\vspace{1cm} {\large Department of Physics, Peking University,
Beijing 100871\\\vspace{.5cm} KITPC, CAS, Beijing 100080, P.R.
China}
\\\vspace{1.5cm}\today\vspace{1cm}
\begin{abstract}
  We construct self-dual string soliton solutions in $AdS_4\times
  S^7$ spacetime, starting from the covariant equations of motion
  of M5-brane. We study the properties of the solutions and find that their action
  are linearized, indicating the BPS nature of the solutions,
  and they have the same electric and magnetic charge. The straight string soliton solution
  represents the configuration of the membranes ending on M5-brane with a straight
  string intersection, and it behaves like the spiky
  solution in flat spacetime. The
  spherical string soliton solution, which could be related to the
  straight one by a conformal transformation, represents the
  membranes ending on M5-brane with a spherical intersection.
\end{abstract}
\end{center}



\newpage


Among the D-brane configurations, the spiky string is one of the
most remarkable. It was first discussed in
\cite{Callan97,Townsend97}. From the DBI action of the D-brane, it
was found that there could exist BPS spiky solution, corresponding
to the configuration of fundamental string ending on the D-brane.
In particular, for the D3-brane in IIB string theory, the spiky
solution could also describe the D1-string or $(p,q)$ string
ending on the D3-brane, corresponding to the magnetic monopole or
dyons in the ${\cal N}=4$ super-Yang-Mills theory. On the other
hand, from dual D-string point of view, this spiky string
configuration could be understood as the D-string
funnel\cite{CMT}, whose noncommutative realization reflects the
non-Abelian nature of the D-strings action.

The spiky solution has a natural generalization in M-theory. The BPS
self-dual string soliton solution constructed in \cite{Howe96} is
its M-theroy cousin. In this case, one has to solve the M5-brane's
equations of motion, involving an interacting 2-form gauge field.
The BPS soliton solution found in \cite{Howe96} represent a
supersymmetric self-dual string on M5-brane, with equal electric and
magnetic charge. The configuration could be interpreted as the
membranes ending on the M5-brane. The dual membrane description of
the configuration was hindered by our ignorance of the non-Abelian
membrane action. Few years ago, Basu and Harvey\cite{Basu} proposed
a generalized Nahm equation and constructed a funnel-like solution
to realize this string soliton configuration. Briefly speaking, the
membranes interact with each other and expand to become an
orthogonal M5-brane. Though it is still an open issue how to derive
the generalized Nahm equation from a non-Abelian membrane action,
the self-dual string soliton solutions turns out to be quite
valuable for us to understand the membrane and M5-brane dynamics in
M-theory. For a nice review of the M-theory branes and their
dynamics, see \cite{Berman07}.

Most of the study on D-branes was focused on the branes in flat
spacetime. In the curved spacetime, it is generically difficult to
find BPS D-brane solution. Of particular interest is the D-branes in
$AdS_5\times S^5$ spacetime, which could be essential to understand
AdS/CFT correspondence. This issue has been addressed in
\cite{SkendrisTaylor}.   Besides the BPS D-brane with fluxes
discussed in \cite{SkendrisTaylor}, it was recently found that there
exists another kind of BPS D-brane which could be taken as the
blow-up of the lower dimensional D-branes. This happens in the
D-brane description of the BPS Wilson
lines\cite{Drukker,Yamaguchi:2006D5,Gomis06}. In this case, for the
Wilson loop operator in the high rank representation, the
corresponding fundamental string blow up to a higher dimensional
D-brane, which could be understood as dielectric
brane\cite{Myers:1999,Rodriguez2006}. The same is true for the
Wilson t'Hooft operators\cite{ChenHe}. In this case, the blow-up of
the D(F)-strings not only involves the interactions among strings
but also involves the interaction between strings and the background
RR-fluxes.

In M-theory, the BPS branes in curved spacetime is even less clear.
Of particular interest are the spacetime $AdS_7 \times S^4$ and
$AdS_4 \times S^7$, both being the maximally supersymmetric
configurations in 11D supergravity. They also arise as the near
horizon geometry of M5-brane and M2-brane supergravity soliton
solution. More importantly they are the playgrounds of AdS/CFT
correspondence in M-theory. The study of the BPS brane configuration
may shed light on the dynamics in M-theory. Unlike the case of
D-branes in $AdS_5\times S^5$, there is no systematic study of the
BPS membrane or M5-brane configurations in these backgrounds, as far
as we know.  The trouble mainly comes from the complicated form of
the equations of motion. In \cite{Chen07}, it has been shown that
there exit M5-brane self-dual string soliton solutions in
$AdS_7\times S^4$ spacetime, corresponding to the M5-brane
description of the Wilson surface operators in high rank
representation in the six-dimensional (2,0) superconformal field
theory. It realizes the picture that the membranes interacting among
themselves and with the background 4-form flux blow up to M5-branes.
The same configuration has also been discussed in \cite{Lunin07}.
However, it is not clear how to understand these configurations from
the dual point of view of the membranes.

In this letter, we construct the self-dual string soliton solutions
in $AdS_4 \times S^7$ background. They are half-BPS but behave
differently from the solutions in \cite{Chen07}. The straight string
soliton solution looks like the spiky solution in the flat
spacetime.


Let us start from  the M5-brane covariant equations of motion in
eleven-dimensions, which was first proposed in \cite{Sezgin97} in
superembedding formalism\cite{Howe96v2}, and was then rederived by
requiring $\k$-symmetry of an open M2-brane ending on the
M5-brane\cite{Chu97}. For other derivations from various actions,
please see \cite{Sezgin99,Sundell97,Sorokin97}. The bosonic
components of the equations include the scalar equation and the
tensor equation. The scalar equation takes the form
 \be\label{scalareq}
 G^{mn}\nabla_m \cE_n^{\underline
 c}=\frac{Q}{\sqrt{-g}}\epsilon^{m_1\cdots
 m_6}\big(\frac{1}{6!}F^{\underline a}_{~m_1\cdots
 m_6}+\frac{1}{(3!)^2}F^{\underline
 a}_{~m_1m_2m_3}H_{m_4m_5m_6}\big)P_{\underline
 a}^{~\underline c}
 \ee
 and the tensor equation is of the form
 \be\label{tensoreq}
 G^{mn}\nabla_mH_{npq}=Q^{-1}(4Y-2(mY+Ym)+mYm)_{pq}.
 \ee
 Here our notation is as follows: the indices from the
beginning(middle) of the alphabet refer to the frame (coordinate)
indices, and the underlined indices refer to the target space
ones.

Let us spend some time to explain the quantities in the above
equations. There exist a self-dual 3-form field strength $h_{mnp}$
on the M5-brane worldvolume. From it, we can define
 \bea
 k_m^{~n}&=&h_{mpq}h^{npq}, \\
 Q&=&1-\frac{2}{3}\Tr k^2, \\
 m_p^{~q}&=&\delta_p^{~q}-2k_p^{~q}, \\
 H_{mnp}&=&4Q^{-1}(1+2k)_m^{~q}h_{qnp}
 \eea
 Note that $h_{mnp}$ is self-dual with respect to worldvolume
 metric but not $H_{mnp}$, which instead satisfies the Bianchi
 identity
 \be
 dH_3=-{\underline F}_4
 \ee
 where ${\underline F}_4$ is the pull-back of the target space 4-form flux. The
 induced metric is simply
 \be
 g_{mn}=\cE_m^{\underline a}\cE_n^{\underline b}\eta_{\underline
 ab}
 \ee
 where
 \be
 \cE_m^{\underline a}=\p_mz^{\underline m}E_{\underline m}^{\underline
 a}.
 \ee
Here $z^{\underline m}$ is the target spacetime coordinate, which
is a function of worldvolume coordinate $z$ through embedding, and
$E_{\underline m}^{\underline
 a}$ is the component of target space vielbein. From the induced
 metric, we can define another tensor
 \be\label{Gmn}
 G^{mn}=(1+\frac{2}{3}k^2)g^{mn}-4k^{mn},
 \ee
 which appear in (\ref{scalareq}). And we also have
 \be
 P_{\underline a}^{~\underline c}=\delta^{\underline
 c}_{\underline a}-\cE_{\underline a}^m\cE_m^{~{\underline c}}.
 \ee
 Moreover, there is a 4-form field strength $F_{{\underline a}_1\cdots {\underline a}_4}$ and its Hodge dual
 7-form field strength $F_{{\underline a}_1\cdots {\underline
 a}_7}$:
 \bea
 F_4&=&dC_3 \nn\\
 F_7&=&dC_6+\frac{1}{2}C_3\w F_4
 \eea
 The frame indices on $F_4$ and $F_7$ in the scalar and the tensor equations have
 been converted to worldvolume indices with factors of
 $\cE_m^{\underline c}$.
  From them, we can define
 \be
 Y_{mn}=[4\star {\underline F}-2(m\star {\underline F}+\star {\underline F}m)+m\star {\underline F}m]_{mn},
 \ee
where \be \star {\underline
F}^{mn}=\frac{1}{4!\sqrt{-g}}\epsilon^{mnpqrs}{\underline F}_{pqrs}
\ee


One of the maximally supersymmetric configurations in
11-dimensional supergravity is $AdS_4\times S^7$ background, which
could be obtained from the near horizon geometry of supergravity
solution of M2-branes. The metric and the bulk 4-form flux take
the form
 \bea
 ds^2&=&\frac{R^2}{y^2}(-dt^2+dx^2+dr^2+dy^2)+4R^2d\O_7^2\\
 F_4&=&-\frac{3R^3}{y^4}dt\w dx\w dr\w dy
 \eea
 where we have rescaled the radius of $AdS_4$ to be $R$ and the
 radius of $S^7$ to be $2R$. Correspondingly the 4-form field strength
 get rescaled by a factor $2$.

 Let the worldvolume coordinates of M5-brane  be
 $\xi_i,i=0,\cdots 5$ and the embedding be
 \bea
 \xi_0=t,~~~ \xi=x,~~~ \xi_2=y,~~~ r=f(y), \\
 \xi_3=\a,~~~ \xi_4=\b, ~~~ \xi_5=\g
 \eea
 where $\a,\b,\g$ are the angular coordinates of a $S^3$ in $S^7$.
 In fact, it turns out that there are freedoms to choose the
 embedding of $S^3$ in $S^7$ since the nontrivial equations of
 motions are actually in the $AdS_4$ part. For simplicity, we take
 the first three angular coordinates in $S^7$. The induced metric
 is then
 \bea
 ds^2_{\mbox{ind}}=\frac{R^2}{y^2}(-dt^2+dx^2+(1+f^{\pr
 2})dy^2)+4R^2(d\a^2+\sin^2\a d\b^2+\sin^2\a\sin^2\b
 d\g^2),\\\label{inducemetric2}
 \eea
where prime denotes the partial derivative with respect to $y$.

The self-dual 3-form field strength could be \be
 h=\frac{a}{2}((2R)^3\sin^2\a\sin\b d\a\w d\b\w
 d\g+\sqrt{1+f^{\pr 2}}(\frac{R}{y})^3dt\w dx\w dy),
 \ee
where $a$ could be a function of $y$. Then we can determine other
quantities: $k^2$, $k_m^{~n}$ and $G^{mn}$:
 \be k^2=k_{mn}k^{mn}=\frac{3}{2}a^4,~~~~~
 k_m^{~n}=\left(\begin{array}{cc}
 -\frac{a^2}{2}I_3&0\\
 0&\frac{a^2}{2}I_3
 \end{array}\right),\ee
 \be G^{tt}=-G^{xx}=-(\frac{y}{R})^2(1+a^2)^2,
~~~~~G^{yy}=(\frac{y}{R})^2\frac{(1+a^2)^2}{1+f^{\pr 2}}, \ee \be
G^{\a\a}=\frac{(1-a^2)^2}{4R^2},~~~G^{\b\b}=\frac{(1-a^2)^2}{4R^2\sin^2\a},~~~
G^{\g\g}=\frac{(1-a^2)^2}{4R^2\sin^2\a\sin^2\b}, \ee where $I_3$
is a rank 3 identity matrix. Also we have
 \be H_3=2a\big(\frac{(2R)^3}{1-a^2}\sin^2\a\sin\b d\a\w d\b\w
 d\g+\frac{\sqrt{1+f^{\pr 2}}}{1+a^2}(\frac{R}{y})^3dt\w dx\w dy\big).
 \ee
 Since the induced field strength of 4-form bulk flux is
 vanishing, we have $dH_3=0$, which requires $a$ to be
 constant.

From the tensor equation of motion (\ref{tensoreq}), we find that
$f$ has to be proportional to $y$, \be f=\k y, \ee with $\k$ being
a constant.  We list the relevant Levi-Civita connection of the
induced metric (\ref{inducemetric2}) in the appendix. Now from the
induced metric
 \be
 ds^2_{\mbox{ind}}=\frac{R^2}{y^2}(-dt^2+dx^2+(1+\k^2)dy^2)+4R^2(d\a^2+\sin^2\a d\b^2+\sin^2\a\sin^2\b
 d\g^2),
 \ee
 we find that M5-brane worldvolume is a
$AdS_3 \times S^3$ spacetime, where $AdS_3$ with radius
$R/\sqrt{1+\k^2}$ is embedded into $AdS_4$ and $S^3$ with radius
$2R$ is completely in $S^7$.

In the scalar equations of motions, since the embedding of $S^3$
part is trivial, we always have $\nabla \cE^{\underline c}_m=0$ for
$S^3$ part. And the right hand side of the equation of motions
indeed vanishes since $P^{\underline c}_{\underline a}=0$ in this
case.

The whole nontrivial part in the scalar equations of motion comes
from the $AdS_3$ part. In this case, we have
 \be
 \cE^{\underline 0}_t=\frac{R}{y}, ~~~ \cE^{\underline 1}_{y}=\frac{R}{y}, ~~~\cE^{\underline
 2}_{x}=\frac{R}{y}, ~~~ \cE^{\underline 3}_{y}=\frac{\k R}{y},
 \ee
 where the vielbein of $AdS_4$ part of the target spacetime are
 \bea
\hat{\th}^0=\frac{R}{y}dt,~~~\hat{\th}^1=\frac{R}{y}dy,~~~\hat{\th}^2=\frac{R}{y}dx,~~~\hat{\th}^3=\frac{R}{y}dr.
\eea The corresponding spin connections could be found in
Appendix.

 It turns out that the scalar equations of motion give only one constraint in this case,
 \be
 \frac{(1+a^2)\k}{\sqrt{1+\k^2}}=2a
 \ee
 or
 \be\label{ak}
 a=\frac{\sqrt{1+\k^2}-1}{\k}
 \ee

Therefore we have a self-dual string soliton solution in
$AdS_4\times S^7$ background once the above relation is satisfied.

It would be interesting to study the properties of this string
soliton solution. Firstly this solution is actually 1/2-BPS. This
could be seen by an appropriate coordinate transformation such
that the above solution is actually the same configuration
discussed in \cite{Yamaguchi03}, where the supersymmetry has been
discussed in detail. The condition (\ref{ak}) is exactly the
condition for keeping half of the supersymmetries. In fact, one
may consider more general background $AdS_4 \times X_7$, where
$X_7$ is a weak $G_2$ manifold. The M5-brane worldvolume take a
form $AdS_3 \times L_3$, where $L_3$ is an associate submanifold
in $X_7$ and $AdS_3$ is embedded in $AdS_4$ as above. Such kind of
M5-brane has been discussed in \cite{Yamaguchi03}, where it has
been shown to be BPS under suitable condition. This kind of
configurations give also self-dual string soliton solutions. One
support for this claim is that the configurations are 1/2-BPS.
Though we are unaware of a direct proof that the BPS
configurations must be the solutions of the equations of motion,
the experience with the BPS Wilson lines and the Wilson surfaces
indicates that this is true. One could also prove the claim by
checking the equations of motion directly. Loosely speaking, since
the $L_3$ embedding is somehow trivial, the nontrivial part is
from the $AdS_3$ part, which leads to the relation (\ref{ak}).

Let us come back to the solution in this paper. The action of this
solution could be worked out. As it is well-known, compared to the
Dp-brane action, which is just a Dirac-Born-Infeld(DBI)-type
action, M5-brane action is much subtler since it describe a
self-interacting chiral 2-form whose field strength is self-dual.
There exist two kinds of action, which are equivalent in the sense
that both of them lead to the same equations of motion. One of
them is so called
PST(Pasti-Sorokin-Tonin)-action\cite{Sorokin9701,Sorokin9711,Schwarz97}.
It is manifestly supercovariant and kappa-invariant, and of a
DBI-like form. It contains an auxiliary scalar, from which the
self-duality condition could be derived as an equation of motion.
This proposal has some troubles in defining a proper partition
function, since the topological class of auxiliary scalar would
break some symmetries of M-theory\cite{Witten96}. The resolution
of this problem is to embed the chiral theory into a non-chiral
one. In \cite{Sundell97}, a nonchiral M5-brane action for
unconstrained 2-form gauge potential has been constructed. In this
action, one has to impose a non-linear self-duality condition to
ensure kappa-symmetry. And the equation of motion for 2-form
potential is equivalent to the Bianchi identity. The action is
given by
 \be\label{nc}
 S=S_{M5}-S_{WZ}=T_5\int (\frac{1}{2}\star {\cal K}-Z_6)
 \ee
 where
 \bea
 {\cal
 K}&=&2\sqrt{1+\frac{1}{12}H^2+\frac{1}{288}(H^2)^2-\frac{1}{96}H_{abc}H^{bcd}H_{def}H^{efa}},\\
  Z_6&=&\underline{C}_6-\frac{1}{2}\underline{C}_3\w H_3,
 \eea
and $T_5$ is the tension of the M5-brane:
 \be
 T_5=\frac{1}{(2\pi)^5l^6_p}.
 \ee
 Here $Z_6$ is the
Wess-Zumino term, in which ${\underline C}_6$ and ${\underline
C}_3$ are the pull-back of the target space gauge potential.
 For the
self-dual soliton solution above, classically two kinds of action
give the same action: \be\label{action}
 S=\int d(\mbox{Vol}) (1+\k^2),
 \ee
where $d(\mbox{Vol})$ is the volume element of M5-brane without
flux. In other words, if we turn off the self-dual field strength
$a=0$, then we have $\k=0$ which implies that we have trivial
embedding of M5-brane to the background. In this case, M5-brane is
$AdS_3\times S^3$ with radius $R$ in $AdS_3$ and radius $2R$ in
$S^3$. Therefore,
 \be
 d(\mbox{Vol})=
(\frac{R}{y})^3(2R)^3\sin^2\a\sin\b dtdxdyd\a d\b d\g.
 \ee
The two terms in the integrand in (\ref{action}) have natural
interpretation: the first term gives the action of M5-brane without
3-form field strength, the second term gives the action of the
membrane ending on the M5-brane. The absence of the square root in
the action indicates that the solution is BPS. This is reminiscent
of the BPS spiky
 solution studied in \cite{Callan97,Townsend97}, which
  has the similar action as above.
This is very similar to the case in flat space time\cite{Howe96}.

Let us check the charges carried by the string soliton. Since our
solution could be taken as M2-branes ending on M5-brane, with
M2-branes' worldvolume extending along $t,x,r$, the charges could
be calculated by
 \bea
 Q_E&=&\frac{1}{{\mbox{Vol}(S^3)}}\int_{S^3} \star H \label{QE}\\
 Q_M&=&\frac{1}{{\mbox{Vol}(S^3)}}\int_{S^3}  H \label{charges}
 \eea
 where $S^3$ is the transverse $S^3$ in $S^7$ and $\star$ here means the Hodge dual
 with respect to the metric of the M5-brane worldvolume without the string
 soliton. The charge of our solution is simply:
 \bea
 Q_M&=&\left(\frac{2R}{l_p}\right)^3\frac{2a}{1-a^2}=\left(\frac{2R}{l_p}\right)^3\k \\
 Q_E&=&\left(\frac{2R}{l_p}\right)^3\frac{2a\sqrt{1+\k^2}}{1+a^2}=\left(\frac{2R}{l_p}\right)^3\k
 \eea
 So our solution has the same electric and magnetic charges. This
 is reminiscent of the self-dual soliton solution in flat
 spacetime\cite{Howe96}. In terms of the charge, one can see the
 action of the membrane of unit charge is the membrane tension
 times its volume.

 It is remarkable that although our solution looks at first sight similar to the
 ones in flat spacetime and the ones corresponding to Wilson
 surface operators in $AdS_7\times S^4$\cite{Chen07}, they are
 really different. Compared to the flat spacetime case, our
 solution depends  on the background 4-form field strength, which
 makes the spacetime curved. And in the Wilson surface case, the
 M5-brane could be taken as the blow-up of the membrane
 interaction and its worldvolume could collapse once the membrane
 charge is turned off. While in the solution we studied in this
 paper, the M5-brane could still make sense even if the membrane
 charge being turned off. This difference is reminiscent of the
 difference between D-brane description of the Wilson(-'t Hooft)
 operators\cite{Drukker,ChenHe} and the BPS brane with fundamental
 fluxes in $AdS_5\times S^5$ spacetime\cite{SkendrisTaylor}.

 Let us consider the perturbation around the string soliton solution.
 For simplicity, we just focus on the one along $\a$ direction.
 The perturbation satisfies the following equation:
 \be
 G^{mn}\nabla_m\p_n\delta=0,
 \ee
where $\delta$ is the perturbation and $G^{mn}$ is the tensor
defined in (\ref{Gmn}). Since $G^{mn}$ is diagonal and one can
separate the $S^3$ part and simplify the above relation to be
 \be
 (-\p_t^2+\p_x^2+\frac{1}{1+\k^2}(\p^2_y-\frac{1}{y}\p_y)+\frac{A}{y^2})\delta=0,
 \ee
 where $A$ is the contribution from $S^3$ part, involving the
 angular momenta quantum numbers. For simplicity, we may take S-wave so
 that $A=0$. The above equation could be simplified more by
setting $\delta=e^{-i\o t+ik_x x}\delta_y$. Then we have
 \be
 (\p^2_y-\frac{1}{y}\p_y+(1+\k^2)B)\d_y=0,
 \ee
where $B=\o^2-k_x^2$. Without losing the generality, setting
$k_x=0$ and defining $\rho=y\o$, we find
 \be
 (\p^2_\rho-\frac{1}{\rho}\p_\rho+(1+\k^2))\d_\rho=0.
 \ee
The solution to this equation is
 \be
 \d_\rho=\rho(A J_1(\sqrt{1+\k^2}\rho)+BN_1(\sqrt{1+\k^2}\rho)),
 \ee
where $J_1$ and $N_1$ are Bessel function and Neumann function
respectively. The asymptotic behavior of the solution could be
discussed straightforwardly. It is remarkable that the
perturbation in our case could be solved exactly, unlike the spiky
string case\cite{Callan97}. Near $\rho\rightarrow 0$ or
$y\rightarrow 0$, the Neumann function blows up so we just choose
the Bessel function above.

The solution we discussed above could be understood as the membranes
ending on the M5-brane with a straight string like intersection. The
other worldvolume direction of the membranes is an infinitely
straight line extending along $r$. Actually one can show that
through a conformal transformation, one can obtain another M5-brane
configuration with spherical intersection. This is more easily seen
in Euclidean signature. Since the embedding of $S^3$ in $S^7$ is
somewhat trivial, our discussion will just focus on the $AdS_4$
part. The metric of Euclideanized $AdS_4$ could be written as
 \be
 ds^2=\frac{R^2}{y^2}(dy^2+dr^2+r^2(d\a^2+\sin^2\a
 d\b^2)).
 \ee
Introducing the coordinates $\eta,\rho$ by the following
relations:
 \bea
 r&=&\frac{R\cos\eta}{\cosh\rho-\sinh\rho},\nn\\
 y&=&\frac{R\sin\eta}{\cosh\rho-\sinh\rho},
 \eea
we can rewrite the metric as
 \be
 ds^2=\frac{R^2}{\sin^2\eta}(d\eta^2+\cos^2\eta(d\a^2+\sin^2\a
 d\b^2)+d\rho^2)
 \ee

 Inspired by the study of the Wilson surface operators in
 \cite{Chen07}, we make the following ansatz
  \be
  \sin\eta=\k^{-1}\sinh\rho,
  \ee
  so the induced metric of the M5-brane worldvolume is
  \be
  ds^2=\frac{R^2}{\sin^2\eta}(\frac{1+\k^2}{1+\k^2\sin^2\eta}d\eta^2+\cos^2\eta(d\a^2+\sin^2\a
 d\b^2)).
 \ee
 It is straightforward to check that the above configuration
 is indeed a solution of the equations of motion provided that $\k$
 satisfied the relation (\ref{ak}). The charges of the solution are
 the same as the straight one. However the action of the solution
 looks more involved. In fact, the three-form gauge potential
 could be taken as
 \be
 C_3=-R^3\frac{\cos^2\eta}{\sin^3\eta}(-\sin\eta+\frac{\k\cos^2\eta}{\sqrt{1+\k^2\sin^2\eta}})
 \sin\a d\eta\w d\a \w d\b.
 \ee
 The nonchiral part of the action is the same as the straight
 case, but the Chern-Simons coupling depends on $C_3$. The action
 consists of two parts:
 one part giving the action of M5-brane without flux,
 and the other part showing the contribution of the string soliton flux. The absence
 of the square root in the action indicates the BPS nature of the
 configuration. The difference of the flux energy with the
 straight one may come from the conformal anomaly discussed in
 \cite{Witten99}.

In this letter, we started from the covariant equation of motion of
M5-brane and constructed the BPS soliton solutions. For the straight
soliton solution we constructed, it is the same configuration
discovered in \cite{Yamaguchi03}, where it has been proved to be
supersymmetric. The same configuration has been discussed in
\cite{Lunin07} from PST action. However, in our work we concerned
more about the physical properties of the configuration. We showed
that its action consists of two parts: one part is the M5-brane
action without flux, the other one is the contribution from the
flux. The action is linearized, indicating its BPS nature. We also
showed that the solution has the same electric and magnetic charge.
It is quite different from the M5-brane self-dual string soliton
solutions discussed in \cite{Chen07}. The essential difference lies
on the fact that the M5-brane configurations in \cite{Chen07} are
the blow-up of the membrane and could collapse once the membrane
charge is turning off, while in the configuration discussed here the
M5-brane is well embedded in the curved background even without
membrane charge. Both the electric and magnetic charges of the
membranes in this configuration are well-defined in the sense of
\cite{Howe96}, while for the configurations in \cite{Chen07}, the
definition of the electric charge is subtle. The solution here looks
similar to the spiky solution discussed in \cite{Callan97}. But in
our case, the construction of the solution involves the background
flux. Furthermore, we showed that through a conformal
transformation, we have a new string soliton solution,  representing
the membranes ending on M5-brane with a spherical intersection.

 One interesting issue is to find an dual description from
non-Abelian membrane interaction\cite{Basu}. However, this seems to
be quite difficult. One obstacle is that we don't know how to
describe the membrane dynamics from the point of view of a
non-Abelian membrane action even in flat spacetime\cite{Baggar}, not
mentioning in curved spacetime. Furthermore, in our background we
have to consider the coupling of membranes with the background
4-form flux. In the D-brane case, the non-Abelian Chern-Simons term
governing the coupling between D-brane and the background RR gauge
potential leads to the dielectric branes\cite{Myers:1999}. In
M-theory, we do not know how to generalize the Chern-Simons coupling
to the non-Abelian case. We wish to come back to this issue in the
future.

\section*{Acknowledgments}

BC would like to thank W. He and L. Zhang for the discussion and
their participation in the initial stage of the project. BC would
like to thank ICTP for its hospitality during his visit. The work
was partially supported by NSFC Grant No. 10535060, 10775002,
NKBRPC (No. 2006CB805905) and the Key Grant Project of Chinese
Ministry of Education (NO. 305001).

\section*{Appendix A: Various Connections}

  In this appendix, we list various connections appeared in our
  calculation. For the induced metric (\ref{inducemetric2}), its Christoffel
symbol
  has nonvanishing components:
 \bea
 \G^{t}_{yt}&=&\G^{x}_{yx}=-\frac{1}{y}\nn\\
  \G^{y}_{tt}&=&-\G^y_{xx}=-\frac{1}{y}\frac{1}{1+f^{\pr 2}}\nn\\
  \G^y_{yy}&=&-\frac{1}{y}+\frac{y^{\pr} y^{{\pr}{\pr}}}{1+y^{\pr
  2}}\nn\\
   \G^\a_{\b\b}&=&-\sin\a\cos\a, \nn\\
   \G^\a_{\g\g}&=&-\sin^2\b\sin\a\cos\a \nn\\
  \G^\b_{\a\b}&=&\frac{\cos\a}{\sin\a},\nn\\
  \G^\b_{\g\g}&=&-\sin\b\cos\b \nn\\
  \G^{\g}_{\a\g}&=&\frac{\cos\a}{\sin\a}, \nn\\
  \G^\g_{\b\g}&=&\frac{\cos \b}{\sin\b}.
  \eea
When $f=\k r$, some of the components above are vanishing.

  For the $AdS_4$ spacetime, its nonvanishing independent components of spin connection
   are
  \bea
  \o^{\underline 1}_{{\underline 0}{\underline 0}}&=&-\frac{1}{R},
  ~~~\o^{\underline 1}_{{\underline 2}{\underline
  2}}=\o^{\underline 1}_{{\underline 3}{\underline
  3}}=\frac{1}{R}
  \eea

\bigskip



\begin{thebibliography}{99}

\bibitem{Callan97}C.G. Callan and J.M. Maldacena, Nucl. Phys. {\bf
B513} (1998)109.

\bibitem{Townsend97}G.W. Gibbons, Nucl. Phys. {\bf B514}
(1998)603.

\bibitem{CMT}N.R. Constable, R.C. Myers and O. Tafjord, {\it
``Noncommutative bion core"}, Phys. Rev. D61 (2000)106009
[hep-th/9911136].

\bibitem{Howe96}P.S. Howe, E. Sezgin and P.C. West, Phys. Lett.
{\bf B399}(1997)49.

\bibitem{Basu}A. Basu and J. Harvey, {\it ``The M2-M5 Brane System and a Generalized Nahm's
Equation"}, Nucl.Phys. B713 (2005) 136-150, [hep-th/0412310].

\bibitem{Berman07}D.S. Berman, {\it ``M-theory branes and their
interactions"}, arXiv: 0710.1707.

\bibitem{SkendrisTaylor}K. Skenderis and M. Taylor, {\it ``Branes in AdS and pp-wave
spacetimes"}, JHEP 0206 (2002) 025.

\bibitem{Drukker}
N.~Drukker and B.~Fiol, {\it ``All-genus calculation of {W}ilson
loops using
  {D}-branes,''}  JHEP {\bf 02} (2005) 010.

\bibitem{Yamaguchi:2006D5}
S.~Yamaguchi, {\it ``Wilson Loops of Anti-symmetric Representation
and D5-branes,''} JHEP 0605 (2006) 037.

\bibitem{Gomis06}J. Gomis and F. Passerini, {\it Holographi Wilson
loops}, JHEP {\bf 0608}, 074 (2006) [hep-th/0604007].

\bibitem{Myers:1999}
R.C. Myers, {\it ``Dielectric-Branes,''}  JHEP {\bf 9912}, 022
(1999) [hep-th/9910053].

\bibitem{Rodriguez2006}D. Rodriguez-Gomez, {\it ``Computing Wilson
lines with dielectric branes"}, Nucl. Phys. {\bf B752} (2006)
316-326 [hep-th/0604031].

\bibitem{ChenHe}B. Chen and W. He, {\it ``1/2 BPS Wilson-'t Hooft
loops"}, Phys. Rev. D{\bf 74}(2006)126008 [hep-th/0607024].

\bibitem{Chen07}B. Chen, W. He, J.B. Wu and L. Zhang, {\it ``M5-branes and Wilson Surfaces
"}, JHEP 0708 (2007)  , [arXiv:0707.3978].



\bibitem{Lunin07}O. Lunin, {\it ``1/2-BPS states in M theory and
defects in the dual CFTs"}, [arXiv:0704.3442].



\bibitem{Sezgin97} P.S. Howe, E. Sezgin and P.C. West, ``Covariant
field equations of the M-theory five-brane", Phys. Lett. {\bf
B399}(1997)49, hep-th/9702008.

\bibitem{Howe96v2}P.S. Howe and E. Sezgin, ``Superbranes", Phys. Lett.
{\bf B390}(1997)133, hep-th/9607227. "$D=11, p=5$", Phys. Lett.
{\bf B394} (1997)62, hep-th/9611008.

\bibitem{Chu97} C.S. Chu and E. Sezgin, ``M-Fivebrane from the open
supermembrane", JHEP {\bf 12}(1997)001, hep-th/9710223.

\bibitem{Sezgin99} E. Sezgin and P. Sundell, ``Aspects of the
M5-brane", hep-th/9902217.

\bibitem{Sundell97}M. Cederwall, B.E.W. Nilsson and P. Sundell,
``An action for the super-5-brane in $D=11$ supergravity",
hep-th/9712059.

\bibitem{Sorokin97}I. Bandos, K. Lechner, A. Nurmagambetov, P. Pasti, D. Sorokin and M. Tonin, {\it ``On the equivalence of
different formulations of the M theory Five-brane"}, Phys. Lett.
{\bf B408} 135-141 [hep-th/9703127].

\bibitem{Yamaguchi03}A. Yamaguchi, ``AdS Branes corresponding to
Superconformal Defects", hep-th/0305007.

\bibitem{Sorokin9701}P. Pasti, D. Sorokin and M. Tonin, {\it
Covariant action for a D=11 five-brane with the chiral field},
Phys. Lett. {\bf B398} (1997)41-46, [hep-th/9701037].

\bibitem{Sorokin9711}I. Bandos, K. Lechner, A. Nurmagambetov, P. Pasti, D. Sorokin and M. Tonin, {\it ``Covariant action for the
superfivebrane of M-theory"}, Phys. Rev. Lett. {\bf
78}(1997)4332-4334, [hep-th/9701149].

\bibitem{Schwarz97}M. Aganagic, J. Park, C. Popescu and J. H.
Schwarz, {\it World-Volume Action of the M Theory Five-Brane},
Nucl.Phys. {\bf B496} (1997) 191-214, [hep-th/9701166].\\
J. Schwarz, {\it Coupling a selfdual tensor to gravity in
six-dimensions}, Phys.Lett. {\bf B395} (1997) 191-195,
[hep-th/9701008].\\
M. Perry and J. Schwarz, {\it  Interacting Chiral Gauge Fields in
Six Dimensions and Born-Infeld Theory}, Nucl.Phys. {\bf B489}
(1997) 47-64, [hep-th/9611065].


\bibitem{Witten96}E. Witten, ``Fivebrane effective action in M
theory", J. Geom. Phys. {\bf 22} (1997)103, hep-th/9610234.






\bibitem{Witten99}R. Graham and E. Witten, {\it ``Conformal anomaly of
submanifold observables in AdS/CFT correspondence"}, Nucl. Phys.
{\bf B546}(1999) 52-64 [hep-th/9901021].


\bibitem{Baggar}J. Bagger and N. Lambert, {\it ``Modeling Multiple
M2's"}, Phys. Rev. D75 (2007) 045020 [hep-th/0611108].








\end{thebibliography}
\end{document}